\documentclass[twocolumn,american,english,manuscript,prl,10pt]{revtex4-1}
\usepackage[T1]{fontenc}
\usepackage[latin9]{inputenc}
\usepackage{amstext}
\usepackage{amsmath,latexsym}
\usepackage{graphicx}
\usepackage{esint}
\usepackage{babel}
\usepackage{hyperref}
\hypersetup{
 hidelinks}
\begin{document}

\title{A new dynamical instability in asymptotically Anti-de-Sitter spacetime}

\preprint{MIT-CTP/4792}

\author{Umut G\"{u}rsoy$^1$, Aron Jansen$^1$ and Wilke van der Schee$^2$}

\affiliation{$^1$Institute for Theoretical Physics and Center for Extreme Matter and Emergent Phenomena, Utrecht University, Leuvenlaan 4, 3584 CE Utrecht, The Netherlands}
\affiliation{$^2$ Center for Theoretical Physics, MIT, Cambridge, MA 02139, USA}

\begin{abstract}
We present fully dynamical solutions to Einstein-scalar theory in
asymptotically  Anti-de-Sitter spacetime with a scalar potential containing particularly
rich physics. Depending on one parameter in the potential we find
an especially interesting regime, which exhibits a dynamically unstable
black brane, already at zero momentum, while nevertheless having
positive specific heat.
We show this using the non-linear dynamics, and give a clear interpretation
in terms of the spectrum of linearized perturbations. Our results translate
directly to their dual strongly coupled non-conformal field theories,
whereby we in particular provide two mechanisms to obtain
equilibration times much larger than the inverse temperature.
\end{abstract}
\maketitle


\noindent {\bf 1. Introduction} Holography has led to crucial qualitative lessons for the
physics of real-time dynamics and equilibration in strongly coupled
quantum theories, such as the fast applicability of hydrodynamics,
within a time of $1/T$, with $T$ being the temperature of the plasma \cite{Chesler:2008hg,Chesler:2010bi,Heller:2011ju,Heller:2012km}.
These studies were mostly performed in a scale invariant setting,
where the $1/T$ scaling is automatic. Much richer physics can be
expected for theories without scale invariance or relativistic symmetry, which is an active
topic of current research both in relation to quark-gluon plasma \cite{Ishii:2015gia,Buchel:2015saa,Fuini:2015hba,Janik:2015waa,Janik:2016btb,Gursoy:2015nza,Du:2015zcb,Attems:2016ugt}
and condensed matter systems \cite{Sybesma:2015oha,Brewer:2015ipa,Gursoy:2016tgf}.
In this Letter we present such a new lesson, where we find stationary phases
which display a dynamical instability
at zero momentum, whereby the time scale of the instability
can be much larger than the inverse temperature.

Einstein-scalar theory in the context of non-conformal holographic QCD was first introduced in \cite{Gursoy:2007cb,Gursoy:2007er} (see also \cite{Csaki:2006ji}). In this Letter we follow Refs. \cite{Gubser:2008yx,Gubser:2008ny}, which introduced a simple potential that could mimic the thermodynamics of QCD as in \cite{Gursoy:2007cb,Gursoy:2007er}, but differs from the latter in the small $\phi$ asymptotics. Our potential contains an exponent $\gamma$ that controls the deviation of the
dual field theory from the conformal limit $\gamma=0$. 
In particular, it was shown in \cite{Gursoy:2008bu} that the dual theory
exhibits a confinement-deconfinement transition for the choice $\gamma\geq 4/3$.
However, most of the detailed analysis focused on the particular case
of $\gamma=4/3$, which is the most realistic choice for the
QCD phenomenology\cite{Gursoy:2009jd}.

In this Letter we initiate a detailed study of the regime $\gamma>4/3$
by solving Einstein-scalar equations of motion both in a time-independent
and a time-dependent setting, to determine respectively the thermodynamics,
and the process of thermalization in the corresponding dual field 
theory. We discover a new critical value  $\gamma_{c}\approx 1.65$
beyond which a new, qualitatively different regime arises. This value depends on the mass of the scalar that corresponds to the scaling dimension of the scalar operator $\Delta$, and the quoted value is for $\Delta=3$.  For smaller $\gamma$ it is known that there is a minimum temperature, at fixed source,
above which there is a  large, stable branch and an unstable branch which has
negative specific heat and a vanishing entropy for large temperature.
For larger $\gamma$  we find that both branches at the minimum temperature have positive specific heat.
Interestingly, the solution with smaller black brane area still has growing entropy for 
large $T$ but is nevertheless dynamically unstable and hence contains an exponentially
growing mode. Lastly, as the temperature goes towards the minimum temperature
both solutions coalesce, whereby the relaxation time diverges.
We illustrate all these features by tuning our initial conditions
such that the evolution first relaxes onto the unstable black brane
solution, after which it decays into the stable black brane.

We hence provide two mechanisms to violate the typical $\mathcal{O}(1/T)$
scaling of the equilibration time. Firstly, as mentioned it is possible to
increase the source to its maximum value, where the equilibration time
diverges. Secondly, at smaller sources it is possible to fine-tune the initial
conditions such that the evolution stays at the unstable time-independent
solution for a time $\gg 1/T$.

\begin{figure*}
\begin{centering}
\includegraphics[width=9cm]{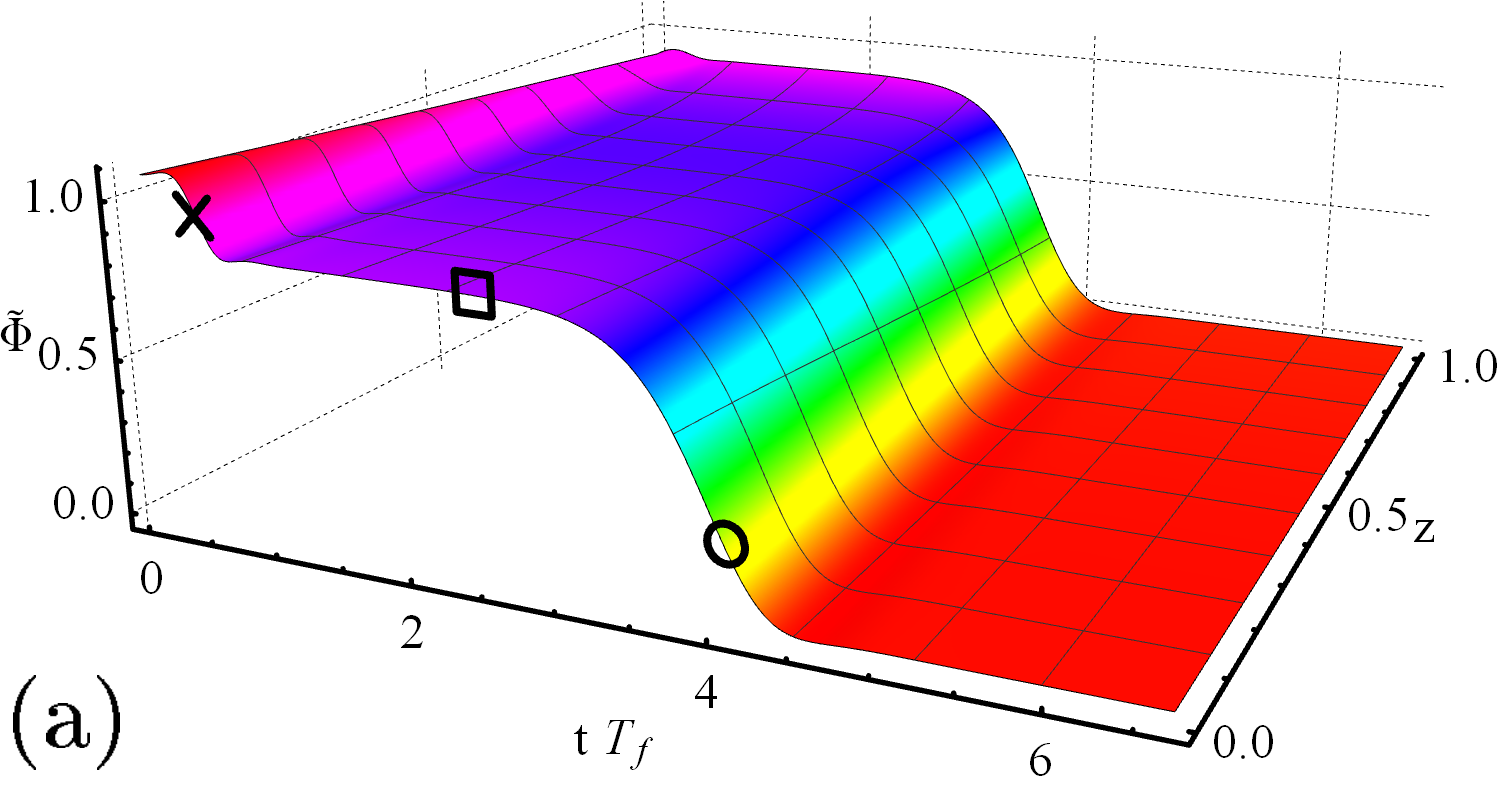}\quad \includegraphics[width=7cm]{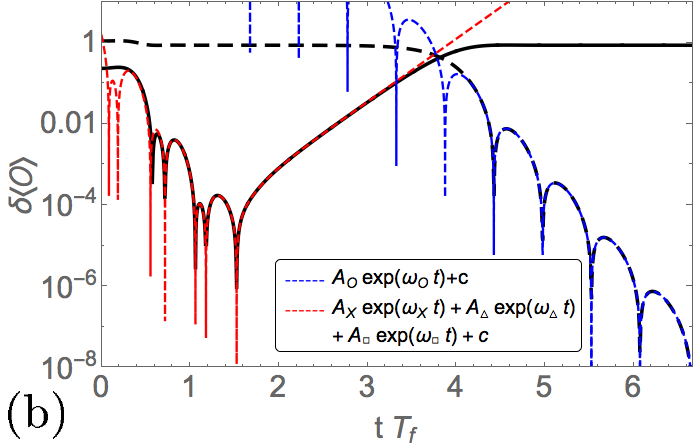}
\par\end{centering}

\caption{(a) We show the dynamical solution with initial conditions given by
$j=0.2$, $a_{4}(0)=-1$ and
$\tilde{\Phi}(z,t)=1.06822$ for the potential (\ref{eq:potential}) with  $\gamma=4\sqrt{2}/3$. 
Interestingly, this evolution
finds both an unstable and a stable solution. The dynamics hence goes through three regimes, as
more clearly shown in (b). There we show the difference $\delta\langle\mathcal{O}\rangle(t)$ between
$\langle\mathcal{O}\rangle (t)$ and the equilibrium values of the unstable and stable phase
in solid and dashed black respectively.
The system first equilibrates to the unstable phase, dominated by two quasi-normal modes (QNM) of
frequencies $(\omega_X, \omega_\Delta)/\pi T = (3.23-1.93 i, -2.18 i)$. After this, the unstable
QNM becomes dominant, and this mode grows exponentially with $\omega_\Box/\pi T =0.92 i$.
Lastly, this mode leaves the linear regime, and the solution finally rings down to the stable solution
with $\omega_O/\pi T =1.82 - 1.79 i$.
\label{fig:unstableplateau}}
\end{figure*}

\begin{figure}
\begin{centering}
\includegraphics[width=7.5cm]{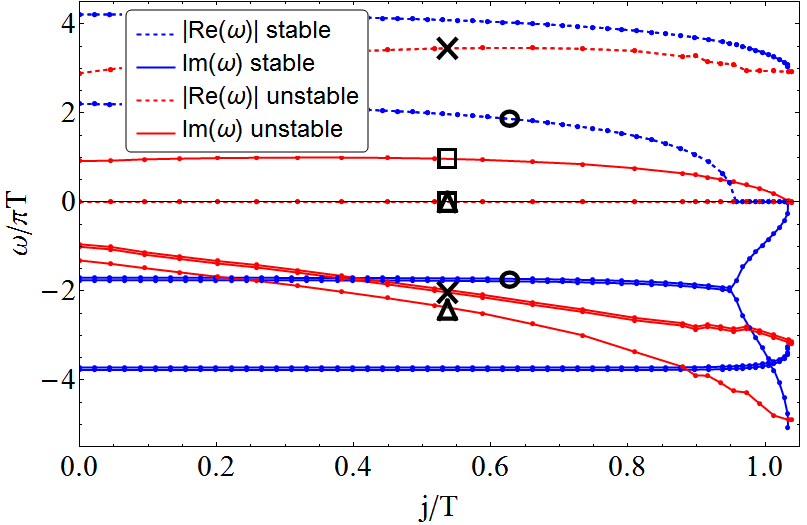}
\par\end{centering}

\caption{We show the first two QNM of both the stable (blue) and unstable (red) solutions as a function of
the source over the temperature. The unstable solution always contains a mode with positive
imaginary part (solid red), which goes to zero for the maximum source allowed.
Interestingly, the lowest QNM of the stable phase has vanishing real part for $j/T>0.95$, after which
the imaginary parts separate. This is needed for the modes to meet at the maximum source,
where both phases coalesce.\label{fig:qnmvsj}}
\end{figure}

Our results also provide an interesting addition to the literature on the
instability of black branes \cite{Konoplya:2011qq}. In particular,
it is different from a Gregory-Laflamme type instability \cite{Gregory:1993vy}.
These type of instabilities are often present in black rings and black
branes in more than four dimensions and involve the break up of the
ring or brane into smaller parts, which can be thermodynamically favourable.
Curiously, the thermodynamics  of black holes crucially relies on
Hawking's computation \cite{Hawking:1974sw}, which requires quantum mechanics.
It is hence not obvious that the classical Einstein equations
would contain a dynamical instability, if a thermodynamic
instability is present (see however \cite{Reall:2001ag}). Nevertheless,
a black brane always contains a dynamical instability
if a thermodynamic instability is present, as conjectured by Gubser and Mitra in \cite{Gubser:2000ec,Gubser:2000mm}
(see also \cite{Hubeny:2002xn,Buchel:2005nt,Buchel:2009bh}) and recently
proved in \cite{Hollands:2012sf}.

Our example is reminiscent of the models studied in \cite{Friess:2005zp}. This reference also has 
hairy black holes in Einstein-scalar theory, which are not unique when specifying all conserved charges.
They note that these non-unique stationary solutions can be unstable to flow towards other solutions.
Since this does not necessarily involve conserved charges this instability can also occur at zero momentum
and with positive specific heat, such as the example presented here in a fully non-linear dynamical setting
\footnote{It is an interesting question if a dynamical instability also implies a thermodynamic instability.
Naively our example is thermodynamically stable with respect to all conserved charges, but as noted
in \cite{Friess:2005zp} it may be more natural to consider a wider definition of thermodynamics which
includes non-conserved charges.}.

\begin{figure*}
\begin{centering}
\includegraphics[width=5cm]{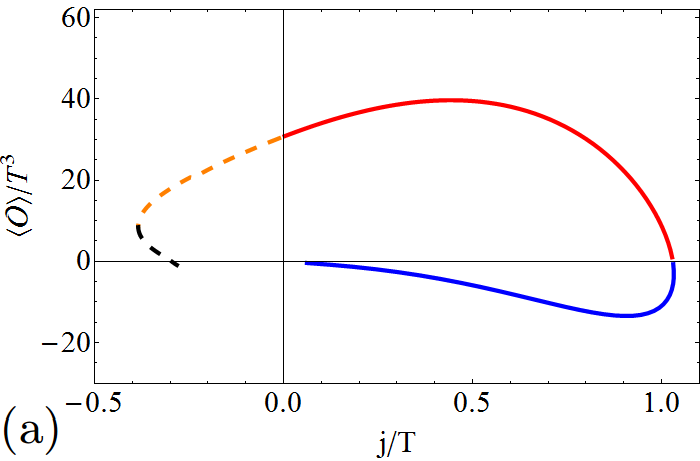}\,\,\includegraphics[width=6cm]{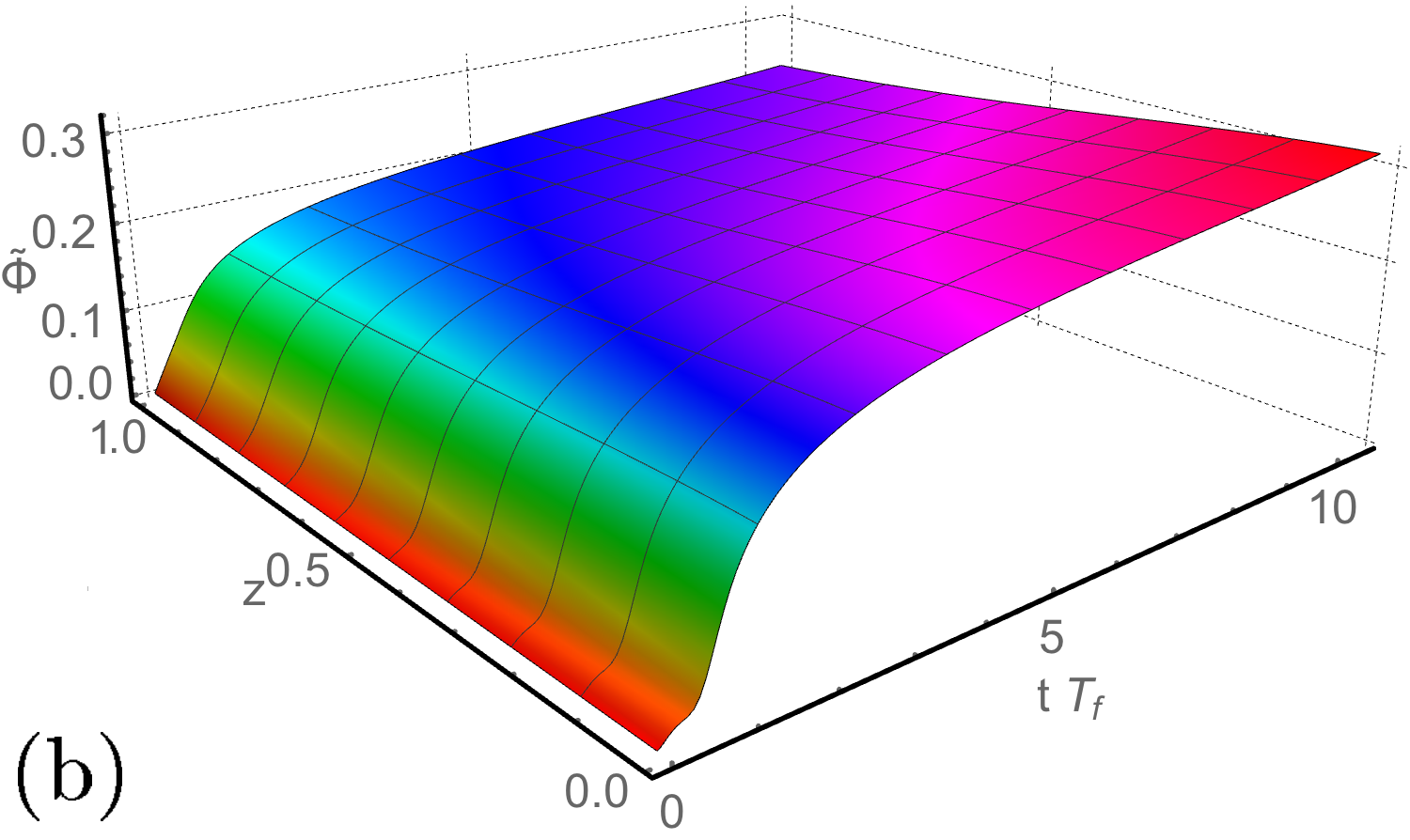}\,\,\includegraphics[width=6cm]{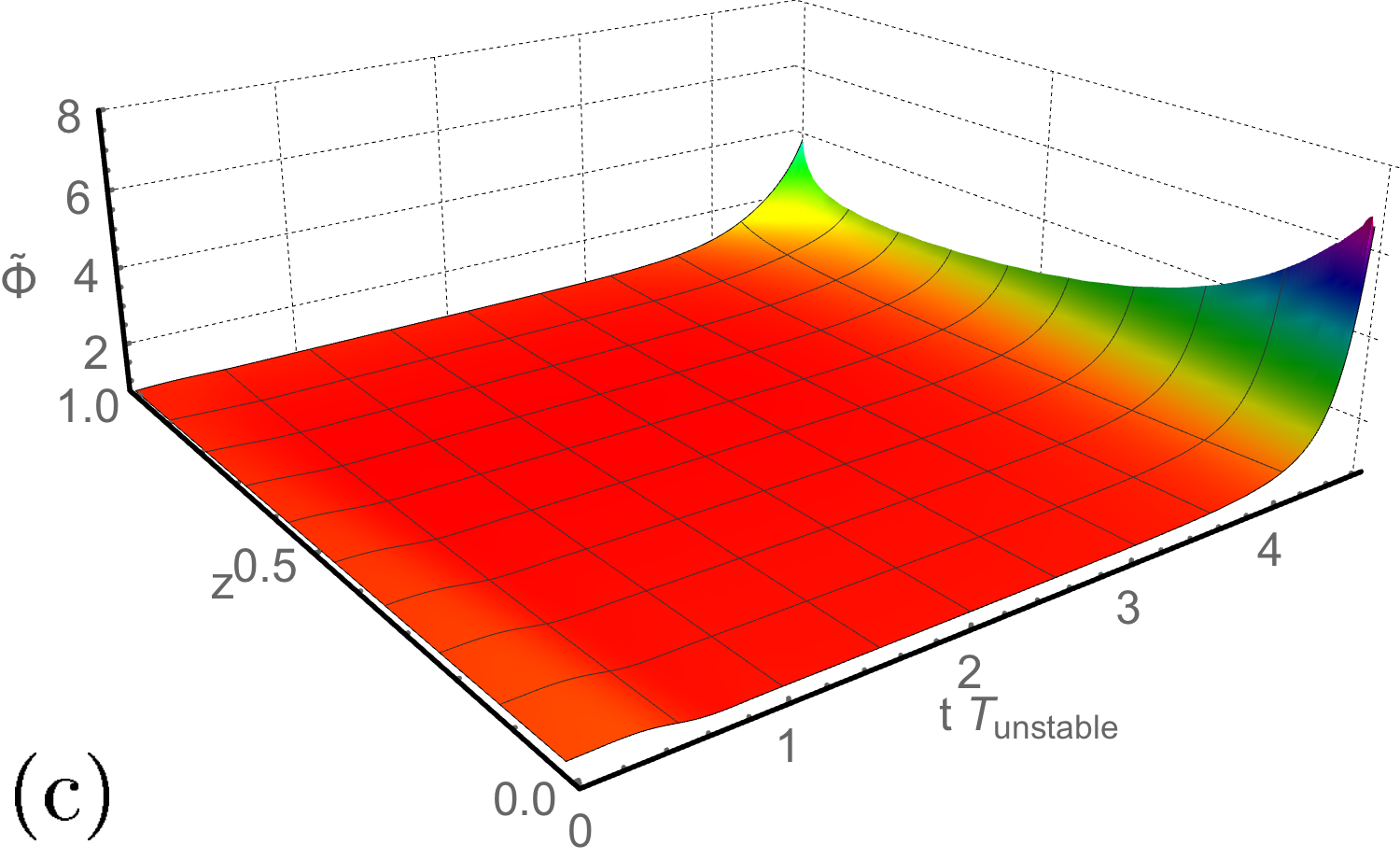}
\par\end{centering}

\caption{
(a) We show the expectation of the scalar operator as a function of the source over the temperature for both
the stable (blue) and unstable (red) phase. There are also two other branches shown dashed, which we comment on in section 3.
(b) For the maximum source the equilibration time diverges, and the system equilibrates in a time $\gg 1/T$.
(c) When evolving initial data slightly different from Fig. \ref{fig:unstableplateau} the coefficient of the
unstable QNM changes sign, and the solution grows instead of decreasing at late times.
\label{fig:blow-up-and-overdamped}}
\end{figure*}

 
\noindent {\bf 2. The model and results} Our confining model is given by the Einstein-scalar action
\begin{equation}
{\cal S}=\frac{1}{16\pi G_{5}}\int d^{d+1}x\sqrt{-g}\left(R-\frac{4}{3}(\partial\Phi)^{2}+V(\Phi)\right),\label{eq:action}
\end{equation}
which includes a scalar field dual to a scalar operator $\mathcal{O}$
with scaling dimension $\Delta$. We follow \cite{Gubser:2008yx,Gubser:2008ny}
and take
\begin{equation}
V(\Phi)=d(d-1)\left(\left(\frac{4\Delta(d-\Delta)}{3d(d-1)}-\frac{\gamma^2}{2}\right)\Phi^2+\cosh(\gamma\Phi)\right),
\label{eq:potential}
\end{equation}
which near $\Phi=0$ reads
\begin{equation}
V(\Phi)=\left(d-d^{2}\right)+\frac{4}{3}\Phi^{2}\left(\Delta^{2}-d\Delta\right)+O\left(\Phi^{4}\right),
\end{equation}
such that we have a negative cosmological constant with $L_{\text{AdS}}=1$,
and the mass of the scalar is independent of the value of the free
parameter $\gamma$. In this Letter we present results for $d=4$
and $\Delta=3$. In order to solve dynamical simulations it is convenient
to use the characteristic formulation of general relativity \cite{Bondi1960Gravitational,Chesler:2008hg,vanderSchee:2014qwa},
which is achieved by using a metric ansatz
\begin{equation}
ds^{2}=2dtdr-Adt^{2}+S^{2}d\vec{x}^{2},\label{metricEF-homogeneous}
\end{equation}
whereby we study homogeneous solutions which only depend on time $t$
and AdS radial coordinate $r$. The equations of motion following
from (\ref{eq:action}) are written down in i.e. \cite{Ishii:2015gia},
and lead to a near-boundary expansion for $\Phi$ as
\begin{equation}
\Phi(r,\,t)=\frac{j}{r}+\frac{j\eta(t)}{r^{2}}+\frac{f_{3}(t)+\left(\frac{3}{8}\gamma^{4}-\frac{4}{9}\right)j^{3}\log(r)}{r^{3}}+O\left(r^{-4}\right),\label{eq:near-boundary-phi}
\end{equation}
where $j$ is a time-independent source for the scalar, $f_{3}(t)$
is the normalizable mode undetermined by the near-boundary analysis
and $\eta(t)$ is a gauge freedom left in (\ref{metricEF-homogeneous}),
which we use to fix the apparent horizon at $r=1$. The normalizable
mode $f_{3}(t)$ has dimension 3 and hence in principle scales as
$j^{3}$ for fixed $j/T$. However, due to the anomalous logarithmic
term in (\ref{eq:near-boundary-phi}) $f_{3}(t)$ also includes a
term $\frac{1}{72}\left(27\gamma^{4}-32\right)j^{3}\log(j)$. The
expectation value of $\mathcal{O}$ can be obtained through holographic
renormalization \cite{deHaro:2000xn,Ishii:2015gia}:
\begin{equation}
\langle\mathcal{O}\rangle=\frac{8f_{3}(t)}{3}+\left(\frac{8}{27}-\frac{\gamma^{4}}{2}\right)j^{3},\label{eq:expectationO}
\end{equation}
where we again notice that this includes an anomalous term proportional
to $j^{3}\log(j)$. As always, there is a scheme dependence in the
renormalization procedure whereby finite counterterms can shift (\ref{eq:expectationO})
without affecting the physics presented.

Lastly, we have to specify initial conditions. Both for numerics and
for presentation it is convenient to treat the near-boundary behavior
of $\Phi$ analytically and hence define $\tilde{\Phi}(z,t)$ as
\begin{eqnarray}
\Phi(z,t) & \equiv & jz+jz^{2}\eta(t)+z^{3}\left(\tilde{\Phi}+\left(\frac{32}{72}-\frac{3}{8}\gamma^{4}\right)j^{3}\log(z)\right)\nonumber \\
 &  & +\frac{3}{8}\left(\frac{32}{9}-3\gamma^{4}\right)j^{3}\eta(t)z^{4}\log(z)
\end{eqnarray}
where $z=1/r$ and by construction $\tilde{\Phi}(0,t)=f_{3}(t)$.
The initial conditions are then fully specified by $\tilde{\Phi}(z,0)$
for $\eta(0)=0$ and the normalizable mode of $A(r,\,t)$, which we
call $a_{4}(0)$.

During the evolution we furthermore keep track of the spectrum of
linearized perturbations around the metric with scalar field, the
so called quasi-normal modes (QNM). We solved for the QNM spectrum
by constructing gauge invariant observables from the linearized equations
\cite{Kovtun:2005ev,Benincasa:2005iv} and used spectral methods to
solve the eigenvalue problem and hence obtain the QNM spectrum \cite{Yaffe:2015MSSTP}.

The main result is captured in Fig. \ref{fig:unstableplateau}(a),
where we take
 $\gamma=4\sqrt{2}/3$
\footnote{We chose this value because this is the maximum value where the spectrum of fluctuations around the vacuum is still well defined \cite{Gursoy:2007er}.} 
, $j=0.2$, $a_{4}(0)=-1$ and
$\tilde{\Phi}(z,t)=1.06822$. Strikingly, the time evolution of this
initial profile goes through three phases. First, within a time of
the inverse temperature, the system relaxes to a time-independent
state. This state is however unstable, and after a time depending
on how close the unstable state was reached the system relaxes again
to the stable solution. We stress that we needed to fine-tune our
initial condition to remain at the unstable state for a long time,
but apart from this fine-tuning this phenomenon is fully generic,
provided $\gamma \gtrsim 1.65$\footnote{In practise this fine-tuning is done by increasing $f_{3}(0)$; for
values higher than the fine-tuned value the evolution looks like Fig.
\ref{fig:blow-up-and-overdamped}(c), which is easy to distinguish
from Fig. \ref{fig:unstableplateau}(a) at late times and hence makes
a binary search algorithm straightforward. Note that we have to fine-tune
the coefficient of one QNM, so we just have to fine-tune one parameter
instead of a function.}. 

In Fig. \ref{fig:unstableplateau}(b) the three regimes are shown
more clearly, first there is the decay to the unstable solution by
the lowest two stable QNM of the unstable phase, which is dominant
since we fine-tuned the coefficient of the unstable QNM to be
small. Then there is a semi-stationary period where this small coefficient
grows exponentially. Finally the solution exits the linear regime around
the unstable solution and rings down in the linear regime
around the stable solution. All three dominant modes are indicated
in Fig. \ref{fig:qnmvsj}, as red QNMs around the unstable
background (first phases) and around the stable background (blue).
Note that the unstable QNM has vanishing real part, which is required
\cite{Konoplya:2011qq}. It is reassuring to see all QNMs of the stable
and unstable solutions meet at $j=j_{\max}$. It is interesting
that the lowest QNM of the stable phase crosses the second QNM to
meet the stable partner of the unstable QNM of the unstable phase.

\begin{figure*}
\begin{centering}
\includegraphics[width=6cm]{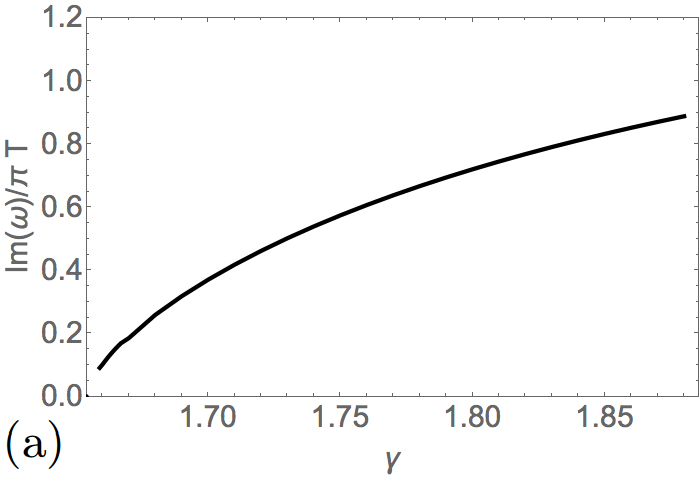}\,\,\includegraphics[width=6cm]{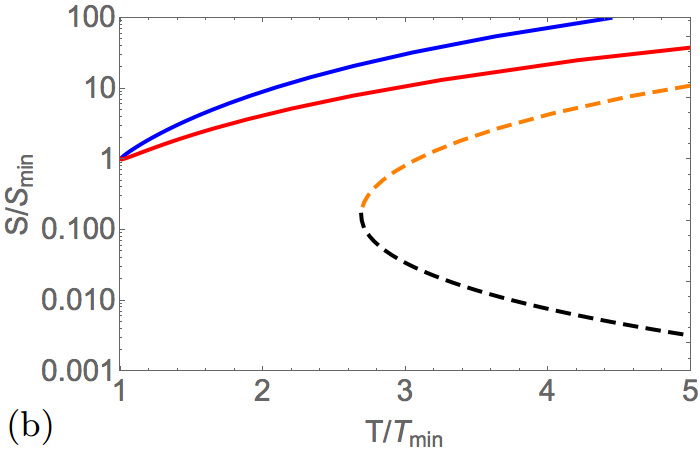}\,\,\includegraphics[width=6cm]{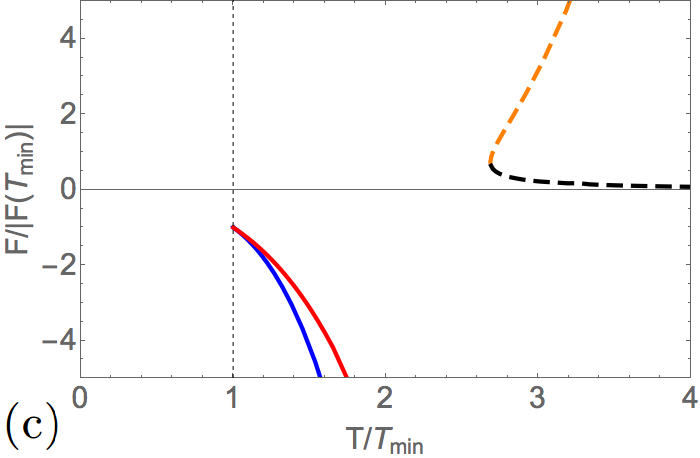}
\par\end{centering}

\caption{
(a) We show the unstable QNM as a function of $\gamma$. It is especially interesting that the relaxation time diverges as $\gamma$ approaches the critical value from above.
(b, c) We show the entropy and free energy as a function of the
temperature over the minimum temperature found for fixed source $j$. Both curves are
normalized with their value at the minimum temperature. Interestingly,
both the dynamically unstable (red) and stable (black) black brane phases have positive specific heat. Third and fourth branches
are shown dashed, which smoothly connect to the thermal gas phase for high temperature, whereby as usual the small branch
has negative specific heat.
\label{fig:vev-and-thermo} }
\end{figure*}

\begin{figure}
\begin{centering}
\includegraphics[width=7.5cm]{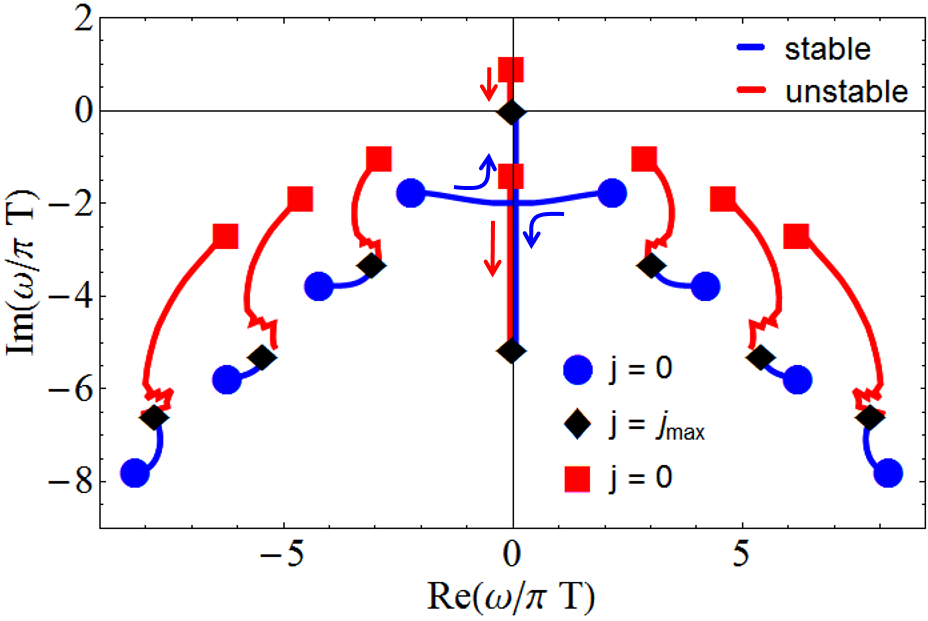}
\par\end{centering}

\caption{Here the first 8 QNMs of Fig. \ref{fig:qnmvsj} are shown
on the complex plane as a function of $j$. The lowest stable QNM (blue)
hits the imaginary axis at $j/T=0.95$, after which the modes separate on
the axis to meet the unstable QNMs at $j=j_\text{max}$.
\label{fig:qnm-all-gamma}}
\end{figure}

Interestingly, for larger source we find the same features, up to a
maximum source $j_{\text{max}}$, as illustrated in Fig. \ref{fig:qnmvsj}
and \ref{fig:blow-up-and-overdamped} (a). At $j=j_{\text{max}}$ the unstable
and stable branch merge, and the lowest QNM of both solutions has $\omega=0$.
This then naturally gives rise to an overdamped relaxation towards
the equilibrium solution, which we show in Fig. \ref{fig:blow-up-and-overdamped}
(b). Clearly, at this point the thermalization takes asymptotically
longer than the standard $1/T$ with $T$ the temperature of the final
state, which is an interesting novelty for non-conformal strongly
coupled systems.

Lastly, we studied the unstable QNM at $j=0$ for different values
of $\gamma$, which is presented in Fig. \ref{fig:vev-and-thermo}(a).
As $\gamma$ approaches the critical value 1.65 from above the imaginary part goes
to zero and we hence again find a diverging relaxation time. We also verified
that for smaller $\gamma$ no instability is present at zero momentum.

We performed several numerical checks. All our simulations satisfy
the constraint equation to better than $10^{-5}$ precision (with the exception of Fig. \ref{fig:blow-up-and-overdamped}(c)). We also
compared the QNM spectrum with fits of the dynamical evolution when
this was in the linear regime, which is a non-trivial check. Lastly,
we used a thermodynamic code to construct the time-independent solution,
which allowed us to study the equilibrium properties of both the stable
and unstable solutions.


\noindent {\bf 3. Discussion} The thermodynamics of the stable and unstable phase is presented in Fig. \ref{fig:vev-and-thermo}(b) and (c), where we
show the entropy and free energy versus the temperature (computed using the method of  \cite{Gursoy:2008za}). Indeed the unstable solution has smaller
entropy than the stable solution, but this phase still consists of a large black hole, in the sense
that the black hole grows as temperature grows. This stands in contrast with two other branches
in our model, shown dashed, which includes a small black hole branch, smoothly 
connecting to the thermal gas phase as the temperature goes to infinity. This
branches are more involved to study dynamically as they are thermodynamically unfavored, 
and we plan to report on this in future work.

As alluded to in the Introduction, one curious feature
is that the  solution shown in Fig. \ref{fig:unstableplateau}(a)
is unstable already at zero momentum, in contrast to the standard Gregory-Laflamme
type instability. This in particular allows for a positive specific heat and  a positive speed of sound squared, as is clear from Fig. \ref{fig:vev-and-thermo}
(c). From a field theory point of view this instability is perhaps less dramatic,
as the plasma need not break up, but evolves homogeneously to a different
phase. Nevertheless, the scalar operator acquires a different expectation value,
and the entropy can at least double for small $j$.

Our results are very generic. For other dimensions, other scaling
dimensions and other potentials with a large enough $\phi^{4}$ contribution
we were able to find the unstable QNM at zero momentum. Also the inclusion
of anisotropy in the metric (\ref{metricEF-homogeneous}) does not give
rise to qualitative changes.

The unstable QNM found in Fig. \ref{fig:unstableplateau} has an interesting
alternative. When choosing $\tilde{\Phi}(z,t)=1.06823$ instead
of $\tilde{\Phi}(z,t)=1.06822$ the coefficient of the unstable QNM
changes sign, and hence $f_{3}(t)$ starts growing exponentially instead
of decreasing. This means the final stable configuration found in
Fig. \ref{fig:unstableplateau} will not be reached. Our numerics
do not allow to determine the final endpoint of this type of evolution,
of which the first moments are displayed in Fig. \ref{fig:qnm-all-gamma}(c).
This would be interesting for future study, whereby we note
that our evolution starts with regular initial conditions, an apparent horizon
and that our model satisfies the null energy condition.

Our results are most easily understood in terms of the QNM spectrum
of both the stable (blue) and unstable solutions (red) on the complex
plane, which we show as a function of $j$ again for $\gamma=4\sqrt{2}/3$
in Fig. \ref{fig:qnm-all-gamma}. Starting at the blue circle for
large black holes with $j=0$ each QNM follows a blue trajectory until
the maximum source $j=j_{\text{max}}$. At this point the lowest QNM
hits the origin and we show the other branch in red, ending again
at $j=0$ with a red square. It is interesting to see the lowest
QNM of the stable phase hits the imaginary axis and hence become overdamped, which is similar to QNMs
in Lifshitz spacetimes \cite{Gursoy:2016tgf}. 

Indeed non-conformal theories contain rich physics. The different
black brane solutions found contain a new dynamical instability
already at zero momentum. This new mechanism leads in
particular to a value of $j/T$ where the quasi-normal mode
frequency goes to zero and hence the relaxation time 
diverges, which we showed to be a generic feature in non-conformal
strongly coupled theories.
\newline

Note added: while finalizing our draft, the paper \cite{Janik:2016btb} appeared on the arXiv that also reports  dynamical instability at vanishing momentum in a similar gravitational setting. However the QNM they find has a small positive imaginary part, about 0.05.


\noindent {\bf Acknowledgments.} We thank Alex Buchel, Steven Gubser, Phil Szepietowski, Takaaki Ishii,
Luis Lehner, Michal Heller, Elias Kiritsis, David Mateos, Francesco Nitti and Chris
Rosen for useful discussions. WS is supported by the U.S. Department
of Energy under grant Contract Number DE-SC0011090. This work was supported by the Netherlands Organisation for Scientific Research (NWO) under VIDI grant 680-47-518, and the Delta-Institute for Theoretical Physics (D-ITP) that is funded by the Dutch Ministry of Education, Culture and Science (OCW).

\bibliographystyle{apsrev4-1}
\bibliography{references}

\end{document}